\documentclass[aps,prl,twocolumn,superscriptaddress,showpacs]{revtex4}

\usepackage{graphicx}

\begin{document}


\title{Splitting of Long-Wavelength Modes of the Fractional Quantum Hall Liquid at $\nu=1/3$}


\author{C.F. Hirjibehedin}\thanks{Present address: IBM Almaden Research Center, San Jose, CA 95120}
\affiliation{Department of Physics, Columbia University, New York,
NY 10027} \affiliation{Bell Labs, Lucent Technologies, Murray Hill,
NJ 07974}

\author{Irene Dujovne}
\affiliation{Department of Appl. Physics and Appl. Mathematics,
Columbia University, New York, NY  10027} \affiliation{Bell Labs,
Lucent Technologies, Murray Hill, NJ 07974}

\author{A. Pinczuk}
\affiliation{Department of Physics, Columbia University, New York,
NY 10027} \affiliation{Bell Labs, Lucent Technologies, Murray Hill,
NJ 07974} \affiliation{Department of Appl. Physics and Appl.
Mathematics, Columbia University, New York, NY  10027}

\author{B.S. Dennis}
\affiliation{Bell Labs, Lucent Technologies, Murray Hill, NJ 07974}

\author{L.N. Pfeiffer}
\affiliation{Bell Labs, Lucent Technologies, Murray Hill, NJ 07974}

\author{K.W. West}
\affiliation{Bell Labs, Lucent Technologies, Murray Hill, NJ 07974}

\date{June 6, 2005}

\begin{abstract}
Resonant inelastic light scattering experiments at $\nu=1/3$ reveal
a novel splitting of the long wavelength modes in the low energy
spectrum of quasiparticle excitations in the charge degree of
freedom. We find a single peak at small wavevectors that splits into
two distinct modes at larger wavevectors. The evidence of
well-defined dispersive behavior at small wavevectors indicates a
coherence of the quantum fluid in the micron length scale. We
evaluate interpretations of long wavelength modes of the electron
liquid.
\end{abstract}



\maketitle


Key properties of the two dimensional (2D) electron liquid phases in
the fractional quantum Hall (FQH) regime are embodied in the lowest
energy neutral excitations in the quasiparticle charge degree of
freedom. These dispersive collective excitations $\Delta(q)$ are
built as a superposition of quasiparticle-quasihole dipole pairs
separated by a distance $x = q l_0^2$
\cite{Lerner&Lozovik1982,Kallin&Halperin1984,Haldane&Rezayi1985,Girvin1985,Girvin1986},
where $q$ is the wavevector and $l_0 = (\hbar c / e B)^{1/2}$ is the
magnetic length for a perpendicular magnetic field $B$. Although the
existence of dispersive modes in FQH systems is assumed in theories
that represent FQH states as quantum fluids, the modes have been
accessible experimentally only at special points in the dispersion.
At the best understood FQH state at Landau level filling factor
$\nu=1/3$, the mode energies have been observed in the long
wavelength limit by inelastic light scattering
\cite{Pinczuk1993,Davies1997,Kang2001}; at the magnetoroton minimum
at $q l_0 \sim 1$ by ballistic phonon studies \cite{Zeitler1999} and
light scattering \cite{Davies1997,Kang2001}; and in the large
wavevector limit by activated magnetotransport
\cite{Willett1988,Du1993,Park1999,Morf2002} and light scattering
\cite{Kang2001}.
\par

Long wavelength excitations of FQH liquids are of major interest.
Such modes manifest the macroscopic length scale of the electron
liquid because the existence of a well-defined dispersion, which
requires wavevector to be a good quantum number, is directly related
to the extent to which the system can be considered translationally
invariant. Studies that map the dispersion in the long wavelength
limit would shed light on the interplay between delocalized states
that extend over macroscopic lengths and states that are localized,
which is of fundamental importance in the understanding of FQH
systems.
\par

Probing the long wavelength dispersion may also clarify an
unresolved issue in the excitation spectrum charge excitations at
$\nu=1/3$. It has been postulated that the lowest energy excitation
at long wavelengths could result from a quadrupole-like excitation
built from two quasiparticle-quasihole pairs, called a two-roton
bound state \cite{Girvin1986,Lee&Zhang1991,He&Platzman1995}. Recent
evaluations \cite{Park&Jain2000,Ghosh&Baskaran2001} have explored
the low-lying charge excitations at $q \rightarrow 0$ by computing
the energy of a pair of bound rotons with opposite wavevector and
found its energy to be below that of the $q \rightarrow 0$ mode
constructed from single quasiparticle-quasihole pairs. Because a
quadrupole-like excitation is predicted to have a qualitatively
different dispersion, experimental determinations of the
long-wavelength dispersion provide a key probe of the character of
the excitations.
\par

We report here the first observation of dispersion in the long
wavelength modes of FQH liquids.  By varying the wavevector $k$
transferred to the 2D system in resonant inelastic light scattering
experiments, we are able for the first time to probe the dispersive
behavior of the lower energy charge excitations at $\nu=1/3$ in the
wavevector range $k l_0 \lesssim 0.15$. In the spectra we find a
mode at the lowest $k$ that splits into two distinct modes at $k l_0
\sim 0.1$. The appearance of two distinct branches in the mode
dispersion reveals the complex nature of the excitation spectrum at
small wavevectors, and is the first direct evidence of the existence
of two different FQH excitations at long wavelengths.
\par

The evidence of dispersive long wavelength excitations provides a
unique measure of the macroscopic length scale of the quantum
liquid. From the changes in the modes for small changes in
wavevector, we are able to infer that coherence in the quantum fluid
occurs in large lakes with macroscopic characteristic lengths that
are in the range of $100 l_0$ and are in the micron length scale. It
is remarkable that behaviors linked to translational invariance
occur on this large length scale given that residual disorder must
exist in the FQH state.
\par


The observed weak dispersion at long wavelengths offers unique
experimental insights into the properties of the FQH liquid and is
consistent with predictions
\cite{Haldane&Rezayi1985,Girvin1985,Girvin1986,Scarola2000a,
He&Platzman1995,Ghosh&Baskaran2001}. Our results indicate that the
long wavelength modes tend to converge as $k \rightarrow 0$,
suggesting the modes may have some mixed character at small
wavevectors \cite{Lee&Zhang1991}. At the largest wavevectors the
higher energy excitation is damped, possibly because of interactions
with a two-roton continuum
\cite{He&Platzman1995,Ghosh&Baskaran2001}.
\par
We present results from two 2D electron systems formed in
asymmetrically doped GaAs single quantum wells. The 2D electron
system in sample A (B) is formed in a $w=250 (330) \text{\AA}$ wide
well and has a density of $9.1 (5.5) \times 10^{10} \text{cm}^{-2}$
with a mobility of $3.6 (7.2) \times 10^6 \text{cm}^2/\text{Vs}$ at
$3.8 (0.33) \text{K}$. Finite well widths weaken interactions in the
2D system. In the simplest approximation this is related to the
ratio $w / l_0$. For samples A and B at $\nu=1/3$ these ratios
differ by approximately $3\%$. Samples are mounted on the cold
finger of a dilution refrigerator with a base temperature of
$45\text{mK}$ that is inserted into the cold bore of a $17\text{T}$
superconducting magnet. The normal to the sample surface is at an
angle $\theta$ from the total applied magnetic field $B_T$, making
$B = B_T \text{cos} \theta$.
\par

\begin{figure}
\includegraphics{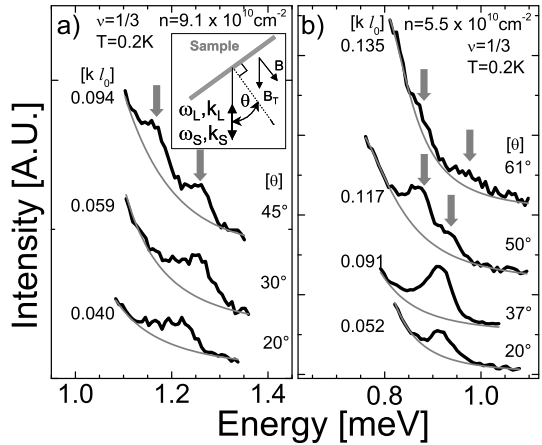}
\caption{\label{fig:figure1} Inelastic light scattering spectra of
low-lying long wavelength charge modes at $\nu = 1/3$ at various
angles $\theta$ in a) sample A and b) sample B. The spectra are also
labelled by the equivalent wavevector $k=(2 \omega_L / c) \text{sin}
\theta$ in units of $1/l_0$. Gray arrows highlight the splitting of
the single peak at small wavevectors into two peaks at larger
wavevectors. Light gray lines show the background. Upper inset in
panel a) shows the inelastic light scattering geometry.}
\end{figure}

Light scattering measurements are performed through windows for
direct optical access, with the laser power density kept below
$10^{-4} \text{W}/\text{cm}^2$. The energy of the incident photons
$\omega_L$ is in resonance with the excitonic optical transitions of
the 2D electron system \cite{Yusa2001,HirjibehedinSSC2003}. Spectra
are obtained in a backscattering geometry, illustrated in the inset
of Fig. \ref{fig:figure1}a. The incident and scattered photons make
an angle $\theta$ with the normal to the sample surface. The
wavevector transferred from the photons to the 2D system is $k = k_L
- k_S = (2 \omega_L / c) \text{sin} \theta$, where $k_{L(S)}$ is the
in-plane component of the wavevector of the incident (scattered)
photon. This allows tuning of the wavevector transferred to the 2D
system by varying the angle of the sample. For a given $\theta$ the
value of $B_T$ is tuned such that $B$ corresponds to $\nu=1/3$.
Sufficient signal can be detected for $\theta < 65^{\circ}$ so that
$k \lesssim 1.5 \times 10^5 \text{cm}^{-1}$. In our samples at $\nu
= 1/3$, this corresponds to $k l_0 \lesssim 0.15$. The uncertainty
in the wavevector of the excitations is primarily set by the finite
solid angle of collection ($\pm 7^{\circ}$ from the vertical) and is
less than $\pm 0.01/l_0$. We therefore present results from spectra
spaced by $\sim 0.02 / l_0$ up to the largest accessible $k$.
\par

\begin{figure}
\includegraphics{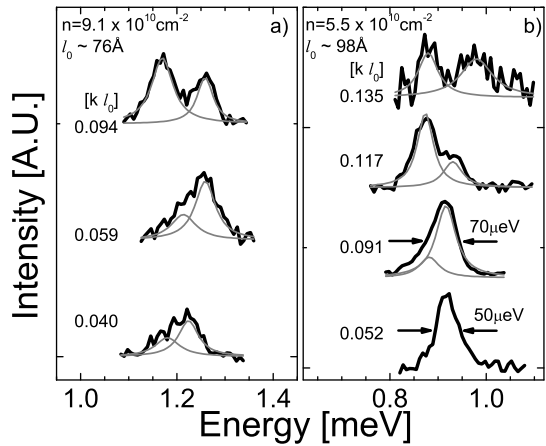}
\caption{\label{fig:figure2} Spectra from Fig. \ref{fig:figure1}a
and \ref{fig:figure1}b with backgrounds subtracted. Gray lines show
fits with two Lorentzian lineshapes.}
\end{figure}

Figure \ref{fig:figure1}a shows light scattering spectra from the
long wavelength charge density excitation at $\nu=1/3$ for various
values of $\theta$ in sample A. The marked angular dependencies
reveal that the wavevector of these excitations is equal to the
wavevector transferred by the photons to the 2D system ($q=k$). We
see that the peak in the spectrum shifts to higher energy for small
changes in wavevector. At the largest wavevector accessible at $1/3$
in this system a remarkable splitting is observed. The two distinct
modes are sharp ($\text{FWHM} < 0.1 \text{meV}$) and are well
separated by $0.1 \text{meV}$. Similar behavior is also observed in
sample B, although as seen in Fig. \ref{fig:figure1}b the energy
scale and the shift in the position of the peak is smaller. Because
smaller values of $B$ are required to access $\nu=1/3$ in sample B,
we are able to probe the excitations at substantially larger $k
l_0$. At the largest wavevector $k l_0 = 0.135$ we see that the
modes are considerably weaker and broader.
\par

The spectra in Fig. \ref{fig:figure1} offer the first direct
evidence for the existence of two distinct FQH modes in the long
wavelength limit at $\nu=1/3$. A similar $k$-dependent splitting
into two branches is not seen in other excitations. For example, the
long wavelength spin wave \cite{Kang2001} remains unchanged
throughout the accessible wavevector range reported here.
\par

The dispersion of the FQH excitations provides a measure for the
length scale of the coherence in the quantum fluid. For the
measurement of a well-defined wavevector dispersion the change in
light scattering wavevector $\delta k$ has to be larger than the
wavevector spacing between modes $2 \pi / L$, where $L$ is a
characteristic length in the fluid. The angular dependence reveals
changes in the modes for $\delta k \cdot l_0 \sim 0.02$. We may
define the lower bound on the macroscopic extent of the
incompressible quantum fluid as $L \gtrsim 2 \pi / \delta k \sim 2
\mu \text{m}$. It is interesting to note that the quantum fluid
manifests the properties of translational invariance on these large
length scales even in the presence of the residual disorder that
must be present in all FQH systems.
\par

\begin{figure}
\includegraphics{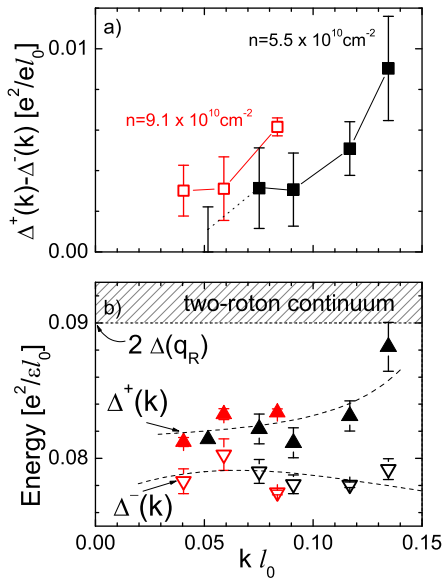}
\caption{\label{fig:figure3} a) Energy difference between the two
modes from samples A (red) and B (black) as a function of $k l_0$
based on fits of spectra. The black bar at $k l_0=0.052$ represents
the uncertainty in the mode separation based on the linewidth of the
observed single peak. b) Energy vs. $k l_0$ for each individual
mode. Open and closed triangles represent energies obtained by fits
with two modes from samples A (red) and B (black). Error bars do not
include the uncertainty in determining the zero-offset ($\lesssim 5
\times 10^{-4} e^2 / \epsilon l_0$) for each spectrum.  The dashed
lines are guides for the eye for possible dispersions of the two
modes.}
\end{figure}

In Fig. \ref{fig:figure2} we show the spectra from Fig.
\ref{fig:figure1} with the backgrounds subtracted. Because two peaks
emerge at larger wavevectors, we interpret the single broadened
peaks at smaller $k$ as unresolved doublets. These modes show
similar dispersive behavior in both samples. To determine the mode
energies, we fit all broadened peaks with two Lorentzian modes and
label the lower (higher) energy mode $\Delta^{-(+)}(k)$. For the
sharpest peak we use the center of the peak with the uncertainty in
the underlying modes estimated by the peak width.
\par

The difference in energy between the peaks is shown in Fig.
\ref{fig:figure3}a as a function of $k l_0$. In both samples the
modes move farther apart at larger wavevectors. This separation is
more pronounced at smaller $k l_0$ in sample A. We interpret these
two modes as two distinct branches of the the long wavelength charge
density excitation for the FQH liquid: a) the long wavelength limit
of the $\Delta(q)$ gap excitation
\cite{Haldane&Rezayi1985,Girvin1985,Girvin1986,Scarola2000a} and b)
a two-roton state at long wavelengths
\cite{Girvin1986,Lee&Zhang1991,He&Platzman1995,Park&Jain2000,Ghosh&Baskaran2001}.
In the long wavelength limit the $\Delta(q)$ mode is predicted to
have a slight downward dispersion while the two-roton mode is
expected to have an upward dispersion.  This is in qualitative
agreement with our observation that the separation between the modes
increases with $k l_0$, though our results do not allow us to
definitively resolve the dispersions of the individual modes.
\par

It is intriguing that the two long wavelength modes seem to converge
as $k \rightarrow 0$. Although $\Delta(q)$ is thought to have a
purely dipolar character near the roton wavevector $q_R$,
calculations suggest that a purely dipolar mode would have
significantly larger energy than a quadrupolar two-roton mode at $k
\rightarrow 0$ \cite{Park&Jain2000,Ghosh&Baskaran2001}. This may
indicate that the $\Delta(q)$ mode has a mixed character at small
wavevectors, becoming purely dipolar only at larger wavevectors near
the roton wavevector $q_R$ \cite{Lee&Zhang1991}.
\par

In Fig. \ref{fig:figure3}b we show the energy vs. $k l_0$ relations
for each individual mode as obtained from the fits, with the energy
in units of $e^2 / \epsilon l_0$ and the wavevector as $k l_0$. We
have scaled down the energies from the higher density sample A by an
additional $4\%$ to account for the difference in finite width
effects, so that the mode energies are consistent as $k \rightarrow
0$. Included in Fig. \ref{fig:figure3}b are possible dispersions for
the two modes that are qualitatively consistent with our results.
The actual dispersion will be much more complex because it should
include interactions between the quasiparticles and interactions
with the continuum of states that is at slightly higher energies
\cite{He&Platzman1995}.
\par



\begin{figure}
\includegraphics{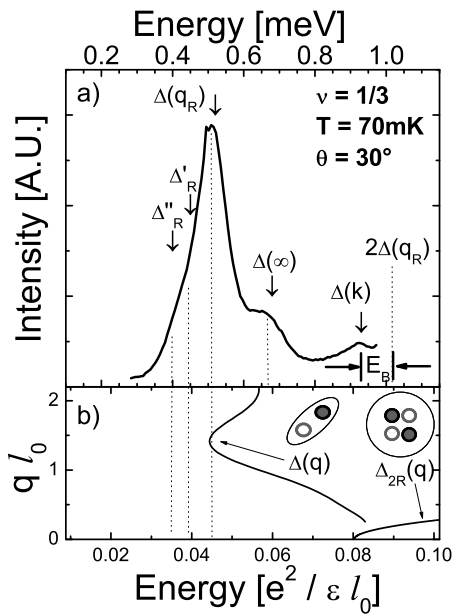}
\caption{\label{fig:figure4} (a) Spectrum at $\nu=1/3$ with
$\theta=30^{\circ}$. Arrows label assigned modes. A number of modes
$\Delta(q_R)$, $\Delta'_R$, and $\Delta''_R$ are seen in the range
of the roton energy. Also indicated is the position of $2
\Delta(q_R)$ and $E_B$. (b) Schematics for the quadrupole-like
$\Delta_{2R}(q)$ dispersion
\cite{He&Platzman1995,Ghosh&Baskaran2001} and the $\Delta(q)$
excitation \cite{Haldane&Rezayi1985,Girvin1985,Scarola2000a}, which
becomes dipole-like at $q_R$.}
\end{figure}

One measure of the two-roton binding energy can be obtained from a
comparison of $\Delta(k) \rightarrow 0$ and the roton energy
$\Delta(q_R)$. In Fig. \ref{fig:figure4} we show a light scattering
spectrum with excitations from critical points in the dispersions at
$\nu = 1/3$. We find that $\Delta(k)$ is somewhat smaller than $2
\Delta(q_R)$. The difference could represent a binding energy $E_B =
0.1 \text{meV}$ that is $10 \%$ of the total two-roton energy. As
seen in Fig. \ref{fig:figure3}, the $\Delta^{+}(k)$ mode approaches
the continuum of two-roton states at the largest $k$. Interactions
between $\Delta^{+}(k)$ and the continuum may be responsible for the
dramatic damping effects seen at the largest $k$ and for the
differences in the mode separations as a function of $k l_0$.
\par

In Fig. \ref{fig:figure4} additional excitations are seen at
energies slightly below the roton energy $\Delta(q_R)$. It is
surprising to find a number of excitations near the roton because
only a single critical point is predicted in the mode dispersion in
that energy range
\cite{Haldane&Rezayi1985,Girvin1985,Girvin1986,He&Platzman1995,Park&Jain2000,Nakajima&Aoki1994}.
These lower energy roton excitations $\Delta'_R$ and $\Delta''_R$ do
not yield a two-roton binding energy when compared to $\Delta(k)
\rightarrow 0$. This suggests that although $\Delta'_R$ and
$\Delta''_R$ may be related to roton excitations they would not
contribute to the formation of a two-roton bound state. It may be
possible for a more complex excitation to be built from a roton and
an additional charged object, such as an additional quasiparticle or
quasihole \cite{Park2001}. Within such an interpretation, the
highest energy roton mode $\Delta(q_R)$ would be the neutral roton
excitation and the lower energy modes $\Delta'_R$ and $\Delta''_R$
would be charged roton excitations. The difference in energy between
the $\Delta(q_R)$ and $\Delta'_R$ modes is $\sim 0.1 \text{meV}$. It
is unlikely that these modes are rotons bound to impurities because
they exists at temperatures well above the $1 \text{K}$ energy
difference (not shown).
\par

Activated magnetotransport gaps \cite{Willett1988,Du1993}, which are
found to be lower in energy than the $\Delta(\infty)$ modes with
which they are normally associated
\cite{Zhang&DasSarma1986,Morf2002}, overlap with the energy of the
roton modes. The neutral roton is normally not expected contribute
to charge transport \cite{Zhang&DasSarma1986}, though the
possibility has been considered \cite{Platzman1985}. However a
charged roton excitation may be able to participate in such
processes. This suggests a possible new role for roton excitations
in magnetotransport in FQH states.
\par

In summary, we present the first experimental evidence of two long
wavelength charge density excitations at $\nu=1/3$ for $k l_0
\lesssim 0.15$. The existence of a dispersion at long wavelengths
implies that the FQH liquid is coherent on length scales in the
micron range. The two observed branches of the long wavelength
excitation spectrum could be associated with the long wavelength gap
excitation $\Delta(q)$ at $\nu=1/3$ that may have dipole-quadrupole
character mixed with a two-roton excitation.
\par



\begin{acknowledgments}
We wish to thank J.K. Jain, S.H. Simon, and H.L. Stormer for helpful
discussions. This work is supported by the National Science
Foundation under Award Number DMR-03-52738, by the Department of
Energy under award DE-AIO2-04ER46133, and by a research grant of the
W.M. Keck Foundation.
\end{acknowledgments}



\end{document}